# The Fe-N system: crystal structure prediction, phase stability, and mechanical properties

Ergen Bao [a,b,&], Jinbin Zhao [a,&], Qiang Gao [a,c], Ijaz Shahid [a], Hui Ma [a,*], Yixiu Luo [a], Peitao Liu [a], Yan Sun [a], Xing-Qiu Chen [a,*] a *Shenyang National Laboratory for Materials Science, Institute of Metal Research, Chinese Academy of Sciences, Shenyang, 110016, China*

b *School of Materials Science and Engineering, Northeastern University, Shenyang, 110819, China*

c *School of Science, Shenyang University of Technology, Shenyang, 110870, China*

**Abstract**:

Nitriding introduces nitrides into the surface of steels, significantly enhancing the surface mechanical properties. By combining the variable composition evolutionary algorithm and first-principles calculations based on density functional theory, 50 thermodynamically stable or metastable Fe-N compounds with various stoichiometric ratios were identified, exhibiting also dynamic and mechanical stability. The mechanical properties of these structures were systematically studied, including the bulk modulus, shear modulus, Young's modulus, Poisson's ratio, Pugh's ratio, Cauchy pressure, Klemen parameters, universal elastic anisotropy, Debye temperature, and Vickers hardness. All identified stable and metastable Fe-N compounds were found in the ductile region, with most exhibiting homogeneous elastic properties and isotropic metallic bonding. As the nitrogen concentration increases, their bulk moduli generally increase as well. The Vickers hardness values of Fe-N compounds range from 3.5 to 10.5 GPa, which are significantly higher than that of pure Fe (2.0 GPa), due to the stronger Fe-N bonds strength. This study provides insights into optimizing and designing Fe-N alloys with tailored mechanical properties.

**Keywords**: First-principles, Fe-N compounds, Mechanical properties, Hardness

## 1. Introduction

Nitriding, a thermochemical treatment process, introduces nitrogen into the surface of metals such as steel, forming a hard and wear-resistant iron nitrides layer. This process significantly enhances surface hardness, wear resistance, and corrosion resistance. As a result, iron nitrides play a crucial role in improving the performance of steel, particularly its surface properties [1, 2]. This is especially important for industries such as automotive, aerospace, and tool manufacturing, where components require enhanced surface hardness, high resistance to scratch and fatigue, and minimal deformation during heat treatment.

Among the various iron nitrides, the most common phases are FeN, ζ-$Fe_2N$, ε-$Fe_3N$, and γ'-$Fe_4N$, which form on the surface of steel during the nitriding process. To investigate their properties, many

---

[&] *These authors contributed equally to this work and should be considered co-first authors.*

[*] *Corresponding authors.*

*E-mail address*: hma@imr.ac.cn (H. Ma), xingqiu.chen@imr.ac.cn (X.-Q. Chen).

related studies have been conducted[3-7]. For example, Weber *et al.* [3] used dynamic nanoindentation to measure the hardness of single-phase nitride γ'-$Fe_4N$ formed by ion-beam synthesis, finding a hardness of 6.6±0.5 GPa, which is 2.5 times higher than that of untreated iron. Tessier *et al.* [4] employed high-temperature solution calorimetry to determine the formation enthalpies of γ''-$FeN_{0.91}$, ζ-$Fe_2N$, ε-$Fe_3N$, and γ'-$Fe_4N$, observing a linear relationship between nitrogen content and formation enthalpy. Takahashi *et al.* [5], combining nanoindentation experiments and *ab initio* calculations, reported the elastic modulus of the iron nitride phase γ'-$Fe_4N$ to be 159±17 GPa. Furthermore, Zhang *et al.* [6] explored the electronic, magnetic, and elastic properties of ε-$Fe_3X$ (X=B, C, N) using first-principles calculations based on density functional theory (DFT), finding that $Fe_3N$ exhibits the highest hardness and magnetic moment among the three phases. Chen *et al.* [7] conducted first-principles calculations to examine the phase stability, magnetism, elasticity, and hardness of binary iron nitrides, demonstrating that nitrogen significantly enhances both the hardness and stability of these phases. However, Gao's hardness formula [8], applied by Chen *et al.* to calculate the hardness of Fe-N alloys, appears to be suitable only for covalent crystals.

Previous studies focused on specific compositions of Fe-N alloys. Expanding the compositional range of Fe-N alloys could provide insight into how varying nitrogen concentrations influence the mechanical properties of these alloys. Additionally, hardness is a critical material property, and most hardness tests utilize the nanoindentation method. However, the preparation of iron nitride samples and the testing procedures can be relatively cumbersome [3, 9]. In this study, crystal structure prediction using USPEX code [10] was employed, which can also identify metastable phases. Based on first-principles calculations, the phase stability and mechanical properties of these Fe-N compounds were investigated. Moreover, hardness was evaluated using semi-empirical models [11]. The aim is to comprehensively verify and complement the structural and mechanical properties of binary iron nitrides. By gaining a more detailed understanding of mechanical properties such as Young's modulus and hardness, this study provides guidance for future research focused on optimizing and designing new Fe-N alloys with specific mechanical properties.

## 2. Calculation details

The evolutionary algorithm from the USPEX 10.5.0 code (Universal Structure Prediction: Evolutionary Xtallography) [10, 12, 13] was employed to search for stable and metastable structures of Fe-N compounds. The structure prediction for Fe-N compounds with varying compositions was performed at 0 K and ambient pressure, with up to 21 atoms per unit cell. The initial population consisted of 100 structures. After structural optimization, 70% of the structures with the lowest enthalpy were selected to generate the next generation (50%—by heredity, 10%—by atomic mutation, 10%—by permutation, 10%—by transmutation, and 20%—by topological random). A total of 100 generations (80 individuals per generation) were calculated, resulting in more than 8400 structures being generated and optimized. During the USPEX calculations, each structure was relaxed in five steps using the Vienna Ab-initio Simulation Package (VASP) [14-16], employing the generalized gradient approximation (GGA) with the Perdew-Burke-Ernzerhof (PBE) functional [17] and the projector augmented wave (PAW) method [18, 19] based on plane waves. The first-order Methfessel-Paxton method [20] was used to determine the electronic occupancy with a smearing of 0.1 eV. For the final relaxation step, the cutoff energy was set to 450 eV, and the energy convergence criterion was $1\times10^{-6}$ eV.

In the calculation of the mechanical properties, the finite difference method in VASP was employed, where each ion was displaced by small positive and negative shifts of ±0.015 Å along the Cartesian directions. The elastic constants were obtained from the first derivative of the stress-strain curve using the VASPKIT code [21]. Homogeneous Monkhorst-Pack *k*-point meshes [22] were employed, with a reciprocal-space resolution of 2π × 0.04 $Å^{-1}$.

The stress $\boldsymbol{\sigma}$ response of solids to an applied strain $\varepsilon$ follows the generalized Hooke's law and can be expressed in Voigt notation [23] $\sigma_i = \sum_{j=1}^{6} C_{ij}\varepsilon_j$, $\varepsilon_i = \sum_{j=1}^{6} S_{ij}\sigma_i$. Here, both strain and stress are represented as vectors with six independent components. $C_{ij}$ is the second-order elastic stiffness tensor, and the compliance tensor $S_{ij}$ is inverse of the stiffness tensor. According to the Hill scheme [24], the effective bulk modulus and shear modulus are arithmetic average of the Voigt and Reuss values, given by $B_{\mathrm{H}} = (B_{\mathrm{V}} + B_{\mathrm{R}})/2$ and $G_{\mathrm{H}} = (G_{\mathrm{V}} + G_{\mathrm{R}})/2$, avoiding extreme estimations due to excessive stiffness or compliance. The Voigt [23] and Reuss [25] bounds on bulk and shear modulus are provided by Eqs. (1)-(4).

$$B_{\mathrm{V}} = \frac{C_{11} + C_{22} + C_{33} + 2(C_{12} + C_{13} + C_{23})}{9} \tag{1}$$

$$G_{\mathrm{V}} = \frac{C_{11} + C_{22} + C_{33} - (C_{12} + C_{13} + C_{23}) + 3(C_{44} + C_{55} + C_{66})}{15} \tag{2}$$

$$B_{\mathrm{R}} = \frac{1}{S_{11} + S_{22} + S_{33} + 2(S_{12} + S_{13} + S_{23})} \tag{3}$$

$$G_{\mathrm{R}} = \frac{15}{4(S_{11} + S_{22} + S_{33}) - 4(S_{12} + S_{13} + S_{23}) + 3(S_{44} + S_{55} + S_{66})} \tag{4}$$

In any case, the Young's modulus ($E$) and the Poisson's ratio ($v$) are given by $E = 9\,BG/(3B + G)$ and $v = (3B - 2G)/(6B + 2G)$. Besides, the elastic anisotropy of crystals can be characterized visually by the universal elastic anisotropic index ($A^{\mathrm{U}}$), which is calculated as $A^{\mathrm{U}} = 5\,G_{\mathrm{V}}/G_{\mathrm{R}} + B_{\mathrm{V}}/B_{\mathrm{R}} - 6$. A value of $A^{\mathrm{U}}$ equal to zero corresponds to elastic isotropy, while a value of $A^{\mathrm{U}}$ equal to 1 indicates the largest anisotropy. Furthermore, Kleinman's parameter ($\zeta$) is used to describe the relative contributions of bond stretching and bond bending in crystal structure deformation [26]. It is calculated as $\zeta = (C_{11} + 8C_{12})/(7C_{11} + 2C_{12})$. Additionally, the Debye temperature ($\Theta_{\mathrm{D}}$) [27] can be estimated from the average sound velocity ($\bar{v}$) using the following equation:

$$\Theta_{\mathrm{D}} = \frac{h}{k_{\mathrm{B}}}\left(\frac{3nN_{\mathrm{A}}\rho}{4\pi M}\right)^{1/3} \cdot \bar{v}\ , \tag{5}$$

where $h$, $k_{\mathrm{B}}$, $M$, $\bar{v}$, $\rho$, $N_{\mathrm{A}}$ and $n$ are the Planck's constant, Boltzmann's constant, molar mass, average wave velocity, density, Avogadro's number, and the number of atoms per formula unit, respectively.

The average wave velocity ($\bar{v}$) can be calculated using $\bar{v} = \left[\left(v_{\mathrm{l}}^{-3} + 2v_{\mathrm{t}}^{-3}\right)/3\right]^{-1/3}$, where $v_{\mathrm{l}}$ and $v_{\mathrm{t}}$ are the longitudinal and transverse wave velocities, respectively, given by $v_{\mathrm{l}} = \sqrt{(3B_{\mathrm{H}} + 4G_{\mathrm{H}})/3\rho}$ and $v_{\mathrm{t}} = \sqrt{G_{\mathrm{H}}/\rho}$.

For the phonon calculations, the structures were expanded into 2×2×2 supercells. The PHONOPY code [28, 29], interfaced with VASP, was used to calculate the force constants via the density functional perturbation theory (DFPT) method, with the $\boldsymbol{k}$-space divided into a 2×2×2 grid using the Monkhorst-Pack method. For the calculation of hardness of Fe-N compounds, Guo's hardness formula [11] was employed. To calculate the electronic density of states (DOS), a reciprocal-space resolution of 2π × 0.02 $Å^{-1}$ was used, and the $\boldsymbol{k}$ space integral was evaluated using tetrahedron method with Blöchl corrections [30]. To calculate Mulliken population, electronic structure calculations were carried out using the CASTEP code within the DFT framework, as implemented in Material Studio [31]. For bond strength analysis, self-consistent calculations were performed with VASP, and the LOBSTER code [32] was used to read the electron wave function and analyze the crystal orbital Hamiltonian population (COHP) [33].

## 3. Results and discussion

### *3.1. Predicted crystal structures*

After 100 generations of USPEX simulations, the calculated convex hull of the Fe-N binary compounds is shown in Fig. 1, which plots the formation energies as a function of nitrogen concentration. The green circles represent the formation energies of the structures searched for each composition, calculated using the following formula:

$$\Delta H = \left(E^{\text{total}}_{\text{Fe}_x\text{N}_y} - xE^{\text{bulk}}_{\text{Fe}} - \frac{y}{2}E_{\text{N}_2}\right)/(x+y). \tag{6}$$

The red solid circles located on the convex hull correspond to the ground-state structures for the composition. In addition to the ground-state structures of Fe and N on both sides of the hull, the remaining two ground-state structures are FeN ($F\bar{4}3m$) and $Fe_3N$ ($P6_322$).

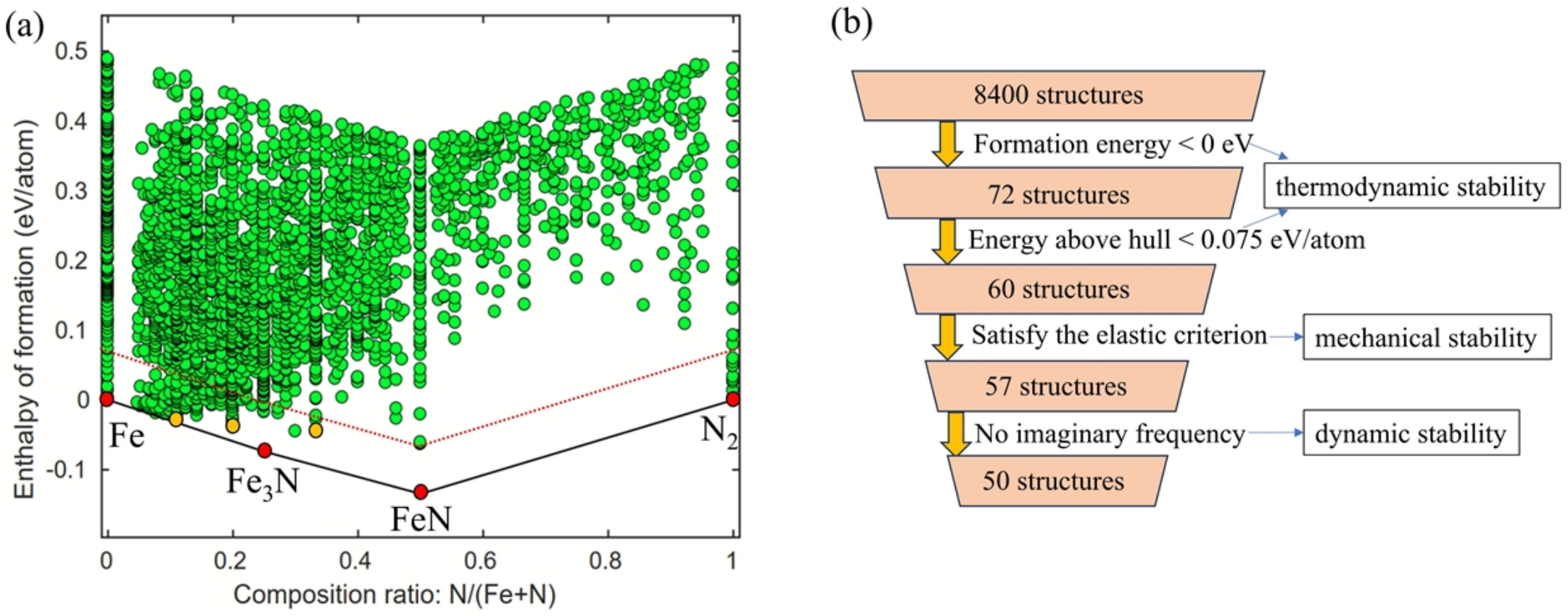

**Fig. 1.** (a) The convex hull of Fe-N binary compounds. (The green circles represent the per-atom formation energy of each compound's searched structure, with red solid circles located on the convex hull representing chemically stable structures and the yellow solid circle represents other structures observed experimentally.) (b) Flow chart for screening thermodynamic, mechanical, and dynamic stable structures of Fe-N compounds.

Among the more than 8400 structures, 72 structures with negative formation energies were selected for further analysis to ensure thermodynamic stability (see Fig. 1b). Additionally, 60 structures with energies above the hull less than 0.075 eV/atom were considered metastable states [34, 35]. Next, using DFT calculations based on the stress-strain method, 57 structures satisfying mechanical stability were identified, with elastic constants meeting the elastic criterion (see supplementary material S2) [36]. Finally, the phonon spectra of 50 of these structures showed no imaginary frequencies, indicating they are dynamically stable (see supplementary material S3). The 50 thermodynamically stable or metastable, dynamically and mechanically stable iron nitrides include $Fe_xN$ ($x$=1~9, 11~16, 19, 20), $Fe_xN_2$ ($x$=9, 13), and $Fe_xN_3$ ($x$=7, 8, 10, 11, 16). Table A1 lists the lattice constants, space groups, formation energies, and energies above hull of these stable and metastable Fe-N compounds. The POSCAR files containing structural information are provided in supplementary material S1.

Interestingly, Fe-N compounds with a nitrogen concentration greater than 0.5 are thermodynamically unstable under normal temperature and pressure. Moreover, DOS calculations show that none of these structures exhibit an energy gap (see supplementary material S4), suggesting they are all metallic. Among these iron nitrides, the lattice constants, formation energies, and energies above hull of FeN ($F\bar{4}3m$), $Fe_2N$ (*Pbcn*), $Fe_3N$ ($P6_322$), $Fe_4N$ ($Pm\bar{3}m$), and $Fe_8N$ (*I*4/*mmm*) are shown in Table 1, compared with experimental values [37-40]. $Fe_{12}N_5$ ($P\bar{3}1m$), which has also been synthesized experimentally, was not found in the calculation due to the maximum atom count being set to 21, whereas the primitive cell of $Fe_{12}N_5$ contains 34 atoms. Most of the remaining 45 metastable structures are predicted for the first time, with phases of $Fe_xN$ ($x$=8~20) showing particularly low energies above hull (<0.02 eV/atom). These phases have not been experimentally observed because they may only stabilize under high-temperature or high-pressure conditions, and thus rapidly transform into more stable phases during experiments. Alternatively, a high nucleation energy barrier, or the presence of impurities and defects in experiments, may also prevent their formation.

**Table 1**

The space groups, calculated lattice constants in conventional cells, calculated formation energies and calculated energies above hull of FeN ($F\bar{4}3m$), ζ-$Fe_2N$, ε-$Fe_3N$, γ'-$Fe_4N$, and $Fe_8N$ (*I*4/*mmm*), compared with the experimental values.

| Phase | Space group | | Lattice constant (Å) | | | Formation energy | Energy above hull |
|---|---|---|---|---|---|---|---|
| | | | *a* | *b* | *c* | (eV/atom) | (eV/atom) |
| FeN | $F\bar{4}3m$ | this work | 4.23 | 4.23 | 4.23 | -0.135 | 0.000 |
| | | Cal. | 4.28 | 4.28 | 4.28 [7] | -0.058 [7] | |
| | | Exp. | 4.31 | 4.31 | 4.31 [37] | -0.060 [4] | |
| ζ-$Fe_2N$ | *Pbcn* | this work | 4.34 | 4.74 | 5.46 | -0.043 | 0.051 |
| | | Cal. | 4.42 | 4.73 | 5.43 [7] | -0.026 [7] | |
| | | Exp. | 4.42 | 4.82 | 5.53 [38] | -0.030 [4] | |
| ε-$Fe_3N$ | $P6_322$ | this work | 4.65 | 4.65 | 4.31 | -0.074 | 0.000 |
| | | Cal. | 4.63 | 4.63 | 4.30 [7] | -0.042 [7] | |
| | | Exp. | 4.70 | 4.70 | 4.38 [39] | -0.052 [4] | |
| γ'-$Fe_4N$ | $Pm\bar{3}m$ | this work | 3.79 | 3.79 | 3.79 | -0.040 | 0.020 |
| | | Cal. | 3.77 | 3.77 | 3.77 [7] | -0.047 [7] | |
| | | Exp. | 3.79 | 3.79 | 3.79 [39] | -0.043 [41] | |
| $Fe_8N$ | *I*4/*mmm* | this work | 5.68 | 5.68 | 6.22 | -0.029 | 0.004 |
| | | Cal. | 5.70 | 5.70 | 6.23 [42] | -0.224 [42] | |
| | | Exp. | 5.72 | 5.72 | 6.29 [40] | - | |

## *3.2. Mechanical properties*

The calculated bulk moduli ($B_H$), shear moduli ($G_H$), Young's moduli ($E$), Poisson's ratios ($v$), Pugh's ratios ($k = B_H/G_H$), Cauchy pressures ($P_C = C_{12}$-$C_{44}$), Kleinman's parameters ($\zeta$), universal elastic anisotropies ($A^U$), and Debye temperatures ($\Theta_D$) of Fe–N compounds are presented in Table A2. The bulk modulus measures a material's resistance to bulk compression. As shown in the Fig. 2a, the bulk modulus generally increases with nitrogen concentration, indicating that compounds become harder to compress as nitrogen content rises, with minimal dependence on precipitate configuration. The calculated bulk modulus of pure Fe is 175.18 GPa, which is consistent with the experimental value of 173±2 GPa [43]. The calculated bulk modulus of γ'-$Fe_4N$ is 190.72 GPa, which is consistent with the experimental value of 196 GPa [5]. Additionally, the shear modulus, defined as the ratio of shear stress to shear strain, is shown in Fig. 3a. Most Fe-N compounds exhibit shear moduli ranging from 50 to 110 GPa. $Fe_8N_3$ (*P*1) and $Fe_2N$ (*Pbcn*) have the highest shear moduli, which are 107.97 GPa and 103.27 GPa, respectively, making them resistant to shear deformation. In contrast, $Fe_4N$ (*Pbcn*) and $Fe_4N$ ($Pm\bar{3}m$) have the lowest shear moduli, at 44.23 GPa and 48.28 GPa, respectively, indicating they are softer and more prone to shear deformation. The calculated shear modulus of γ'-$Fe_4N$ is 48.28 GPa, close to the experimental value of 59 GPa [5].

Poisson's ratio is the ratio of the transverse (lateral) contraction strain to the longitudinal (axial) extension strain when the material is stretched. As shown in Fig. 2b, most Fe-N compounds exhibit predominantly metallic bonding. The typical Poisson's ratio for covalently bonded crystals is around 0.1, while for ionic and metallic compounds, it is typically around 0.25 and greater than 0.33 [9], respectively. As seen in Fig. 2c, all Pugh's Ratio ($k=B/G$) are greater than 1.75, indicating that all stable and metastable Fe-N compounds lie in the ductile region [44]. These compounds undergo significant deformation before fracture, making them more flexible and capable of withstanding stress without breaking. According to Fig. 2d, the Cauchy Pressure ($P_C=C_{12}$-$C_{44}$) of most Fe-N compounds is greater than 0 GPa, suggesting that these compounds possess isotropic metal-like bonds, consistent with the Poisson's ratio observation. However, a few Fe-N compounds ($Fe_4N$-*Cmcm*, $Fe_9N$-*C*2/*m*, $Fe_{15}N$-*C*2/*m*) exhibit more directional covalent bonding characteristics [45], and their hardness is generally higher than that of other metal-bonded compounds.

As shown in the Fig. 2e, most Fe-N compounds exhibit uniform elastic properties in all directions. However, several phases of $Fe_4N$ (*Pbcn*, *Fmmm*, *Fdd*2, *P*1, and *Pmna*) display strong elastic anisotropy, with universal elastic anisotropy values of 10.11, 3.07, 3.01, 2.65, and 2.60, respectively. This anisotropy arises from the low symmetry of their lattice and the directional nature of chemical bonds [46]. Furthermore, for most Fe-N compounds (especially those with N concentration≥0.2), Kleinman's parameters exceed 0.5, indicating that, under strain, changes in bond length are more pronounced than changes in bond angle. For a few Fe-N compounds with Kleinman's parameter less than 0.5, the system tends to adjust the bond angle rather than stretch or compress the bond length.

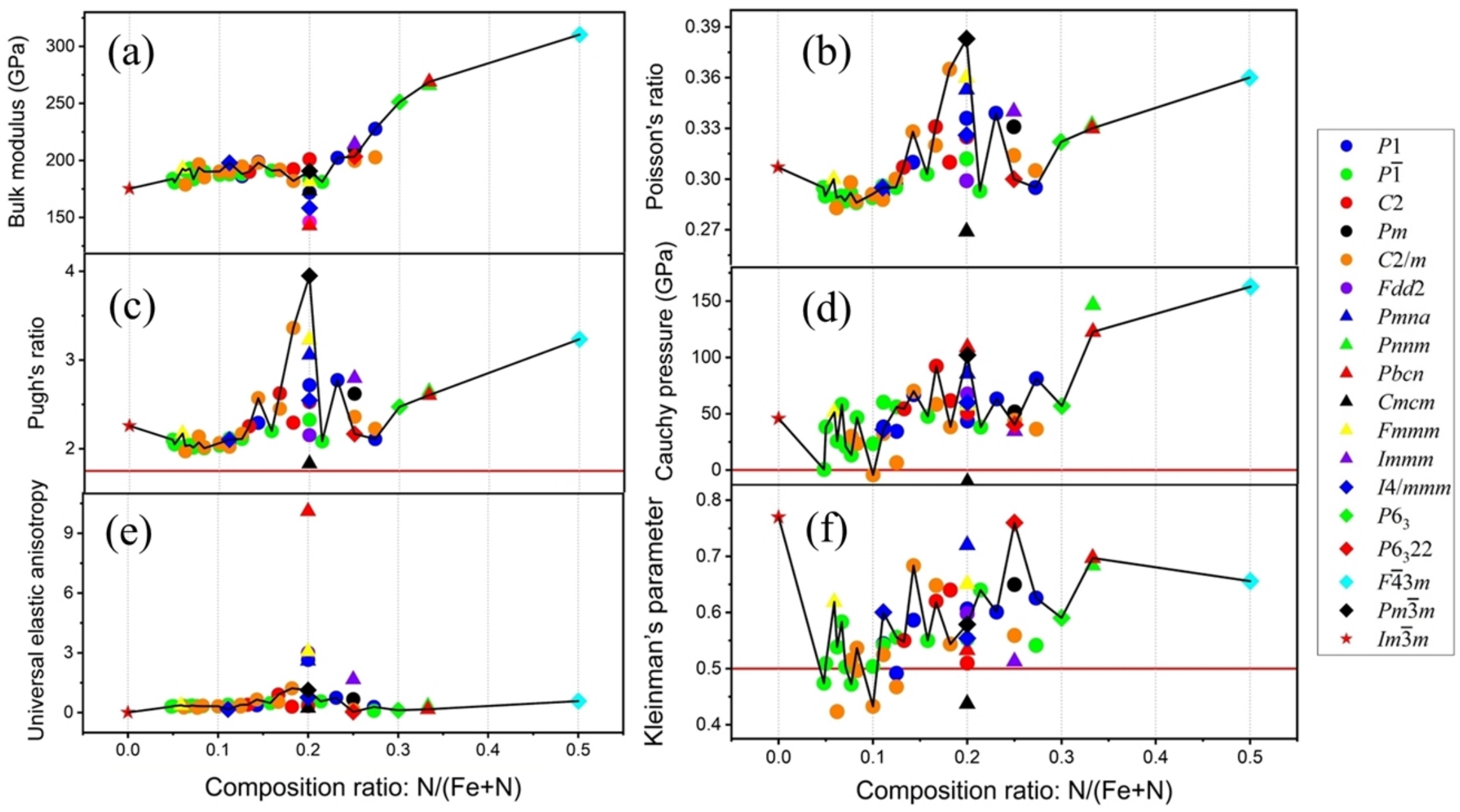

**Fig. 2.** The calculated (a) bulk moduli, (b) Poisson's ratios, (c) Pugh's ratios, (d) Cauchy pressures, (e) universal elastic anisotropies, and (f) Kleinman's parameters of stable and metastable Fe-N compounds predicted by USPEX. The black line indicates the line between compositions with the lowest energy above hull.

Due to the dominant influence of the shear modulus on the calculations for Young's modulus and Debye temperature, the trends of Young's modulus and Debye temperature as functions of nitrogen concentration exhibit a certain similarity to that of shear modulus. Young's modulus is defined as the ratio of tensile stress to tensile strain in a material, quantifying its response to stretching or compression under a given load, and it represents the material's stiffness. Meanwhile, the magnitude of Young's modulus reflects the strength of interatomic bonding. As shown in Fig. 3b, most Fe-N compounds have a Young's modulus in the range of 150 to 280 GPa. $Fe_8N_3$ ($P1$), $Fe_2N$ ($Pbcn$), $Fe_7N_3$ ($P6_3$), and $Fe_2N$ ($Pnnm$) exhibit the highest Young's modulus, with values of 279.72, 274.65, 268.72, and 267.85 GPa, respectively, indicating their high stiffness and strong chemical bonding. In contrast, $Fe_4N$ ($Pbcn$), $Fe_4N$ ($Pm\bar{3}m$), $Fe_9N_2$ ($C2/m$), and $Fe_4N$ ($Fmmm$) have the lowest Young's modulus, with values of 120.27, 133.57, 147.88, and 152.58 GPa, respectively.

The Debye temperature is a characteristic temperature that correlates with the highest vibrational frequency of a crystal lattice. Materials with high Debye temperatures generally exhibit higher thermal conductivity. As shown in Fig. 3c, FeN ($F\bar{4}3m$), $Fe_2N$ ($Pbcn$), and $Fe_8N_3$ ($P1$) exhibit the highest thermal conductivities, while $Fe_4N$ ($Pbcn$), $Fe_4N$ ($Pm\bar{3}m$), and $Fe_9N_2$ ($C2/m$) exhibit the lowest. The thermal conductivities of experimentally synthesized Fe-N compounds are ranked as follows: FeN ($F\bar{4}3m$) > $Fe_2N$ ($Pbcn$) > $Fe_3N$ ($P6_322$) > $Fe_8N$ ($I4/mmm$) > $Fe_4N$ ($Pm\bar{3}m$).

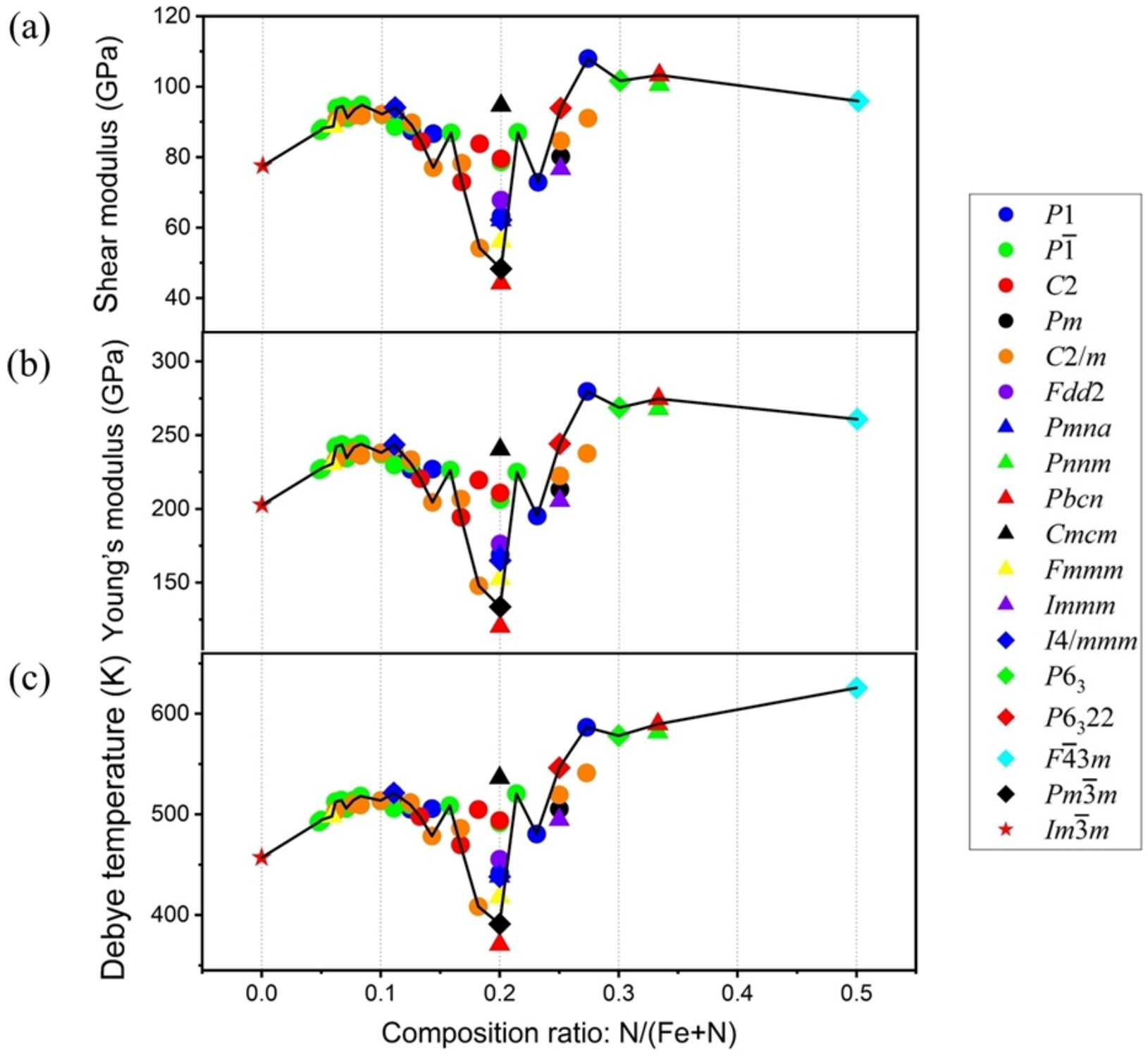

**Fig. 3.** The calculated (a) shear moduli, (b) Young's moduli, and (c) Debye temperatures of stable and metastable Fe-N compounds predicted by USPEX. The black line indicates the line between the compositions with the lowest energy above hull.

### *3.3. Vickers hardness*

Vickers hardness is determined by the pressure applied to the unit area of an indentation surface and may depend more on the bonding properties than on traditional microstructural characteristics [47]. Accurately calculating hardness has long been recognized as a complex task, with many research groups making noteworthy contributions in this field [8, 11, 48-52]. Teter *et al.* [52] established a strong correlation between hardness and shear modulus, expressed as $H_{\mathrm{V}} = 0.151 G_{\mathrm{H}}$. Gao *et al.* [48] linked hardness to electronic structure and proposed a semi-empirical method for evaluating the hardness of covalent and polar covalent crystals. Subsequently, Gao *et al.* [8] introduced a model for calculating hardness based on Mulliken population analysis, emphasizing the role of bond strength and electron density in determining hardness for covalent crystals. Chen *et al.* [49] presented a macroscopic hardness model incorporating the Pugh modulus ratio, $H_{\mathrm{V}} = 2 (k^2 G_{\mathrm{H}})^{0.585} - 3$, which showed good agreement with experimental data. Zhao *et al.* [50] expanded Chen's dataset and proposed a simplified model, $H_{\mathrm{V}} = 0.16 k^{0.5} G_{\mathrm{H}}$, using machine learning.

Guo *et al.* [11] developed a formula incorporating metallicity, bond ionicity, and covalent character, considering the influence of metal composition and *d* valence electrons in transition metal carbides and nitrides. According to Guo's model, Vickers hardness can be calculated using the equation:

$$H_{\mathrm{V}} = A_0\, n_e^{2/3}\, d^{-2.5}\, \mathrm{e}^{-1.191 f_i - 32.2 f_{\mathrm{m}}^{0.55}} \,, \tag{7}$$

where $A_0$ is equal to 350 for bonds without *d* valence electrons, $n_e (= \sum_i n_i Z_i / V)$ is the valence electron density, $f_i$ is the Phillips ionicity of the chemical bond, described by $f_i =$ $\left[1-\exp\left(-\frac{|p_C-p|}{p}\right)\right]^{0.735}$ [51], $p$ and $d$ are the overlap population and the bond length, respectively, $p_C$ is the overlap population of diamond (0.75), and $f_m$ is the metallicity parameter, described by $f_m = 0.026D_F/N_e$, where $D_F$ and $N_e(=\sum_i n_i Z_i)$ are the total DOS value at the Fermi level and the number of the valence electrons in the cell, respectively. Since pure iron does not contain covalent bonds, the Mulliken overlap population between atoms cannot be evaluated.

To compare the accuracy of different models, we calculated the Vickers hardness of α-Fe, ε-$Fe_3N$, γ'-$Fe_4N$ and $Fe_8N$ (*I*4/*mmm*) using different hardness models and compared the results with the experimental values shown in Table 2. As evident from Table 2, Guo's microscopic model accurately predicts the hardness of Fe-N compounds, with small mean absolute errors (MAEs) and root mean square errors (RMSEs) for three compounds. Chen's and Zhao's macroscopic hardness models also provided good predictions, with MAEs around 4 GPa. However, since the datasets used to construct Chen's and Zhao's models did not involve metallic-bonded crystals [49, 50], their models are not suitable for α-Fe. This indicates that the relationship between Vickers hardness and the power of shear modulus in pure metals is not purely linear.

**Table 2**

The calculated Vickers hardness of α-Fe, ε-$Fe_3N$, γ'-$Fe_4N$ and $Fe_8N$ (*I*4/*mmm*) using different hardness models, compared with experimental Vickers hardness, as well as the corresponding MAEs and RMSEs.

| Phase | Space group | $H_V$ (GPa) (this work) | | | | $H_V$ (GPa) (Exp.) |
|---|---|---|---|---|---|---|
| | | Teter's model | Chen's model | Zhao's model | Guo's model | |
| α-Fe | $Im\bar{3}m$ | 11.7 | 6.8 | 8.3 | - | 2.0 [53] |
| ε-$Fe_3N$ | $P6_322$ | 14.2 | 8.5 | 10.2 | 5.7 | 5.9 [54] |
| γ'-$Fe_4N$ | $Pm\bar{3}m$ | 7.3 | 0.9 | 3.9 | 9.9 | 6.6 [9] |
| $Fe_8N$ | *I*4/*mmm* | 14.2 | 8.9 | 10.4 | 5.5 | 6.9 [55] |
| MAS | | 6.5 | 3.8 | 4.2 | 1.6 | |
| RMSE | | 7.4 | 4.1 | 4.4 | 2.1 | |

The Fe-N bond in iron nitrides exhibits covalent, ionic and metallic characteristics. Therefore, Guo's model was used to calculate the Vickers hardness of all stable and metastable Fe-N phases, as presented in Table A3. Fig. 4a shows that the Vickers hardness values of Fe-N compounds range from 3.5 to 10.5 GPa, significantly higher than the experimental value of pure Fe (2.0 GPa) [53], which is consistent with the known phenomenon that the hardness of Fe surfaces increases after nitriding [53, 56]. Among the Fe-N compounds, FeN ($F\bar{4}3m$), $Fe_4N$ ($Pm\bar{3}m$), $Fe_{16}N$ (*Fmmm*), and $Fe_3N$ (*Immm*) have the highest predicted Vickers hardness values (10.17, 9.94, 7.89, and 7.44 GPa, respectively) due to their short bond lengths, large overlap populations, and small DOSs at the Fermi level. In contrast, $Fe_7N_3$ ($P6_3$), $Fe_4N$ (*Pmna*), $Fe_9N_2$ (*C*2/*m*), and $Fe_4N$ (*Fmmm*) exhibit the lowest predicted hardness values (3.59, 3.71 and 4.08 GPa, respectively).

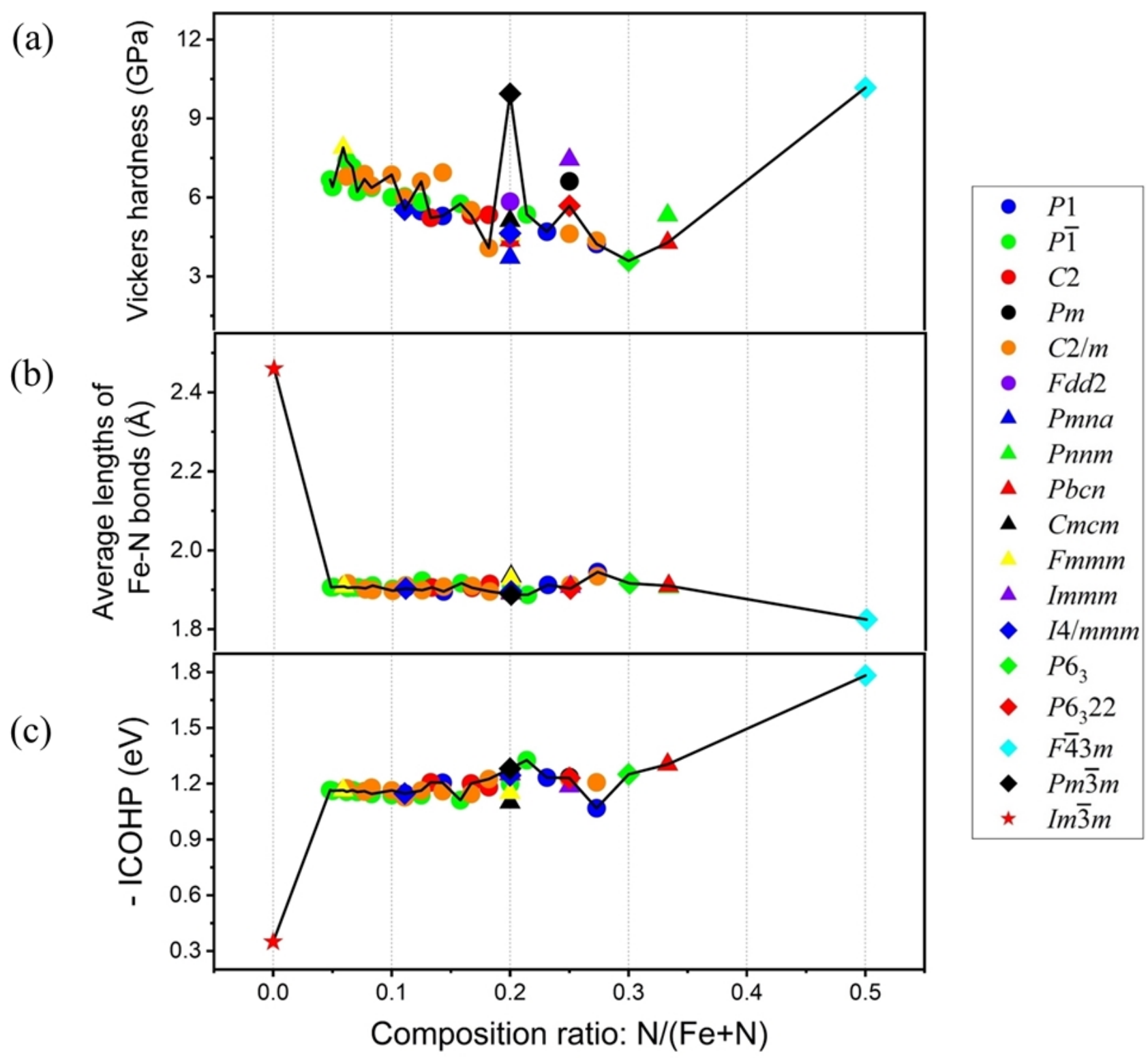

**Fig. 4.** (a) The calculated Vickers hardness using Guo's model, (b) average lengths of Fe-N bond, and (c) -ICOHPs of Fe-N bonds for stable and metastable Fe-N compounds predicted by USPEX. The black line indicates the line between the compositions with the lowest energy above hull. (For α-Fe, (b) and (c) correspond to the length and -ICOHP of Fe-Fe bonds.)

The distributions of electron localization function (ELF) [57] were calculated to visualize the electron distribution, as shown in Fig. 5. It is observed that local electrons accumulate near the N atom in FeN, $Fe_4N$ and $Fe_{16}N$, indicating that Fe-N forms ion bonds in these compounds, which contributes to their higher hardness. In contrast, the electrons in $Fe_7N_3$ are more dispersed, suggesting that Fe-N forms covalent bonds in $Fe_7N_3$, leading to lower hardness compared to other Fe-N compounds.

To quantitatively assess bond strength, the COHPs [33] of Fe-N bonds were calculated. The energy integral of the COHP (ICOHP) quantifies the contribution of a specific bond to the band energy. A higher ICOHP indicates stronger bond strength. As shown in Figs. 4b and 4c, FeN exhibits the strongest bonds, with shorter bond lengths compared to other Fe-N compounds. Moreover, the bond strengths of Fe-N compounds are greater than that of pure Fe, and their bond lengths are shorter. Clearly, high hardness is associated with strong chemical bonds and short bond lengths.

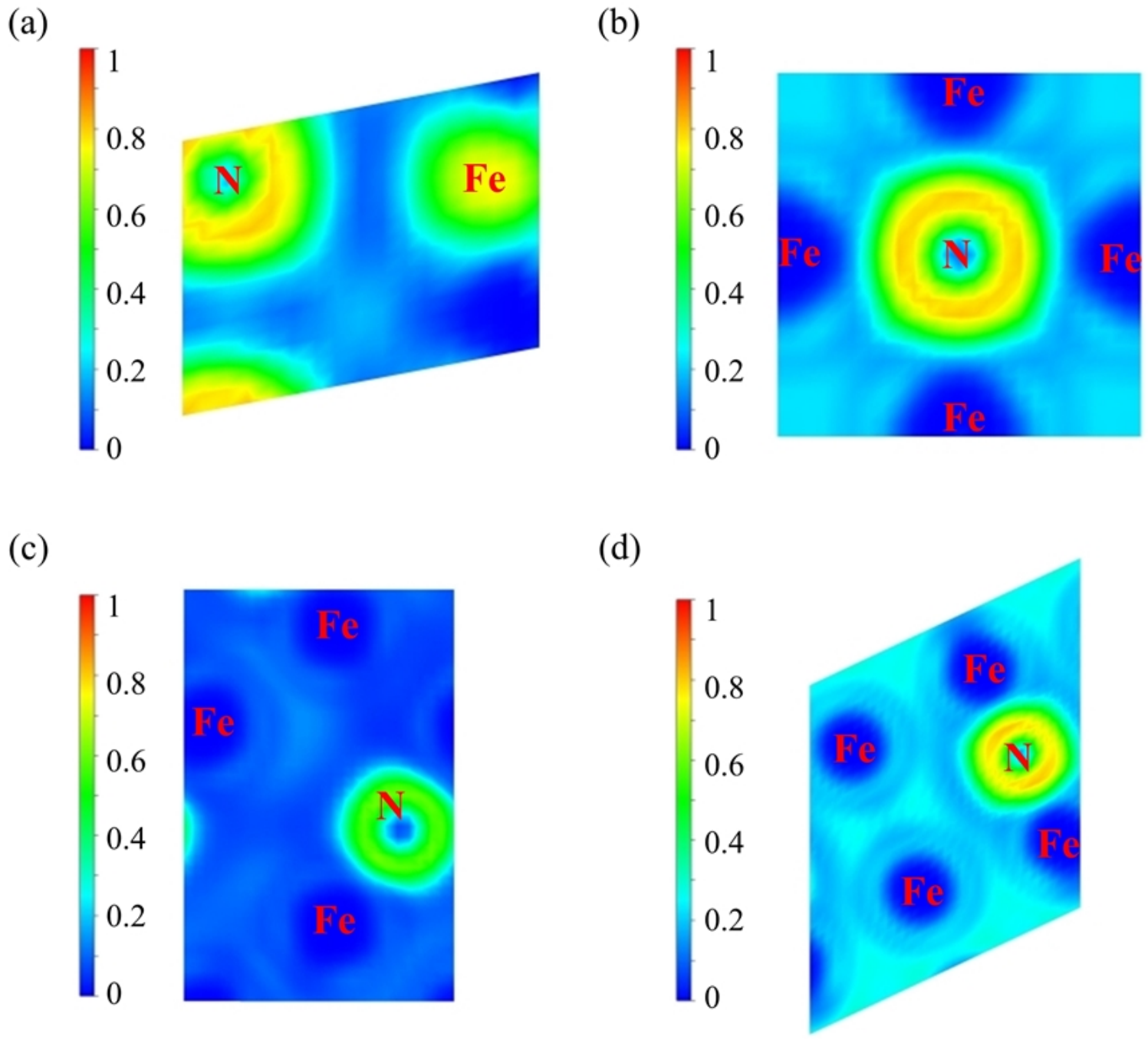

**Fig. 5.** ELF slice in (a) (0 2 $\bar{1}$) plane for FeN, (b) (0 0 1) plane for $Fe_4N$, (c) (4.55 $\bar{1}$ 0) plane for $Fe_7N_3$, and (d) (0 1 0) plane for $Fe_{16}N$.

## 4. Conclusions

By combining the variable composition evolutionary algorithm with first-principles calculations based on DFT, 2 stable and 48 metastable iron nitride phases were identified, and their mechanical properties were systematically investigated. It has been found that that all the stable and metastable Fe-N compounds lie within the ductile region, with most exhibiting uniform elastic properties and isotropic metallic bonding. Additionally, the bulk modulus generally increases as the nitrogen concentration increases. The shear moduli of most Fe-N compounds range from 50 to 110 GPa, with $Fe_8N_3$ (*P*1) and $Fe_2N$ (*Pbcn*) having the highest shear moduli at 107.97 and 103.27 GPa, respectively. The Young's moduli of most Fe-N compounds fall between 150 and 280 GPa, with $Fe_8N_3$ (*P*1), $Fe_2N$ (*Pbcn*), $Fe_7N_3$ ($P6_3$), and $Fe_2N$ (*Pnnm*) exhibiting the highest values of 279.72, 274.65, 268.72, and 267.85 GPa. In terms of thermal conductivity, the Fe-N compounds follow this ranking: FeN ($F\bar{4}3m$) > $Fe_2N$ (*Pbcn*) > $Fe_3N$ ($P6_322$) > $Fe_8N$ (*I*4/*mmm*) > $Fe_4N$ ($Pm\bar{3}m$). Moreover, the calculated Vickers hardness ranges from 3.5 to 10.5 GPa, significantly exceeding the experimental value of 2.0 GPa for α-Fe. Among them, FeN ($F\bar{4}3m$), $Fe_4N$ ($Pm\bar{3}m$), $Fe_{16}N$ (*Fmmm*), and $Fe_3N$ (*Immm*) exhibit the highest predicted Vickers hardness values of 10.17, 9.94, 7.89, and 7.44 GPa, respectively, due to their short bond lengths and strong bond strengths.

### CRediT authorship contribution statement

**Ergen Bao:** Conceptualization, Data curation, Formal analysis, Investigation, Methodology, Validation, Visualization, Writing – original draft, Writing – review and editing, **Jinbin Zhao:** Data curation, Formal analysis, Investigation, Validation, **Qiang Gao:** Conceptualization, Formal analy sis, Methodology, Writing – original draft, Writing – review and editing, **Ijaz Shahid:** Data curation, Formal analysis, **Hui Ma:** Conceptualization, Formal analysis, Investigation, Methodology, Project administration, Writing – original draft, Writing – review and editing, **Yixiu Luo:** Software, Visualization, **Peitao Liu:** Writing – review & editing, **Yan Sun:** Writing – review & editing, **Xing-Qiu Chen:** Funding acquisition, Project administration, Software, Supervision, Writing – review and editing.

## Declaration of Competing Interest

The authors declare that they have no known competing financial interests or personal relationships that could have appeared to influence the work reported in this paper.

## Data availability

The data that support the findings of this study are available from the corresponding author upon request.

## Acknowledgements

This work was supported by National Science and Technology Major Project of China (No. J2019-VI-0019-0134), Liaoning Provincial Natural Science Foundation Project of China (No. 2023-MS-017), National Natural Science Foundation of China (No. 52188101) and Special Projects of the Central Government in Guidance of Local Science and Technology Development (2024010859-JH6/1006).

## Supplementary materials

Supplementary material associated with this article can be found, in the online version, at doi:xxx.

**Appendix A.** Summaries of crystal structure parameters and mechanical properties of stable and metastable Fe-N compounds predicted by USPEX.

**Table A1:** A summary of the phases, compositions, lattice constants in the conventional cells, space groups, formation energies, and energies above hull of stable and metastable Fe-N compounds predicted by USPEX.

| Phase | Composition N/(Fe+N) | Space group | | Lattice constant (Å) | | | Formation energy | Energy above hull |
|---|---|---|---|---|---|---|---|---|
| | | symbol | number | *a* | *b* | *c* | (eV/atom) | (eV/atom) |
| FeN | 0.5 | $F\bar{4}3m$ | 216 | 4.23 | 4.23 | 4.23 | -0.135 | 0.000 |
| $Fe_2N$ | 0.333 | *Pbcn* | 60 | 4.34 | 4.74 | 5.46 | -0.043 | 0.051 |
| | 0.333 | *Pnnm* | 58 | 4.31 | 4.70 | 2.77 | -0.028 | 0.066 |
| $Fe_7N_3$ | 0.3 | $P6_3$ | 173 | 7.21 | 7.21 | 4.30 | -0.045 | 0.041 |
| $Fe_8N_3$ | 0.273 | *P*1 | 1 | 4.27 | 4.93 | 5.35 | -0.019 | 0.060 |
| | 0.273 | *C*2/*m* | 12 | 9.25 | 5.41 | 4.62 | -0.012 | 0.068 |
| $Fe_3N$ | 0.25 | $P6_322$ | 182 | 4.65 | 4.65 | 4.31 | -0.074 | 0.000 |
| | 0.25 | *Pm* | 6 | 4.44 | 2.81 | 6.82 | -0.008 | 0.066 |
| | 0.25 | *C*2/*m* | 12 | 8.35 | 3.82 | 5.30 | -0.018 | 0.056 |
| | 0.25 | *Immm* | 71 | 2.68 | 3.77 | 7.96 | -0.009 | 0.065 |
| $Fe_{10}N_3$ | 0.231 | *P*1 | 1 | 4.46 | 5.18 | 6.31 | -0.008 | 0.060 |
| $Fe_{11}N_3$ | 0.214 | $P\bar{1}$ | 2 | 5.08 | 5.15 | 6.71 | 0.000 | 0.063 |
| $Fe_4N$ | 0.2 | $Pm\bar{3}m$ | 221 | 3.79 | 3.79 | 3.79 | -0.040 | 0.020 |
| | 0.2 | $I4/mmm$ | 139 | 3.77 | 3.77 | 7.35 | -0.025 | 0.034 |
| | 0.2 | $Pmna$ | 53 | 5.32 | 4.55 | 4.27 | -0.031 | 0.028 |
| | 0.2 | $P\bar{1}$ | 2 | 4.70 | 5.24 | 6.67 | -0.016 | 0.043 |
| | 0.2 | $P1$ | 1 | 4.57 | 5.04 | 5.25 | -0.029 | 0.030 |
| | 0.2 | $Pbcn$ | 60 | 4.29 | 5.31 | 9.08 | -0.025 | 0.034 |
| | 0.2 | $Fmmm$ | 69 | 5.56 | 6.88 | 11.03 | -0.004 | 0.055 |
| | 0.2 | $C2$ | 5 | 8.91 | 4.54 | 5.20 | -0.033 | 0.026 |
| | 0.2 | $Cmcm$ | 63 | 7.88 | 6.76 | 3.92 | -0.007 | 0.052 |
| | 0.2 | $Fdd2$ | 43 | 8.63 | 9.10 | 5.27 | -0.036 | 0.023 |
| $Fe_9N_2$ | 0.182 | $C2$ | 5 | 12.05 | 4.34 | 4.70 | -0.001 | 0.053 |
| | 0.182 | $C2/m$ | 12 | 8.41 | 3.83 | 7.79 | -0.008 | 0.046 |
| $Fe_5N$ | 0.167 | $C2/m$ | 12 | 7.73 | 3.92 | 5.36 | -0.007 | 0.042 |
| | 0.167 | $C2$ | 5 | 11.99 | 4.31 | 5.08 | -0.008 | 0.042 |
| $Fe_{16}N_3$ | 0.158 | $P\bar{1}$ | 2 | 6.16 | 6.36 | 6.52 | 0.000 | 0.046 |
| $Fe_6N$ | 0.143 | $P1$ | 1 | 5.10 | 5.12 | 6.70 | -0.014 | 0.028 |
| | 0.143 | $C2/m$ | 12 | 7.62 | 3.94 | 5.12 | -0.008 | 0.035 |
| $Fe_{13}N_2$ | 0.133 | $C2$ | 5 | 11.95 | 4.26 | 7.31 | 0.000 | 0.039 |
| $Fe_7N$ | 0.125 | $C2/m$ | 12 | 8.57 | 3.96 | 5.18 | -0.012 | 0.025 |
| | 0.125 | $P1$ | 1 | 5.07 | 6.33 | 6.41 | -0.002 | 0.035 |
| | 0.125 | $P\bar{1}$ | 2 | 5.07 | 6.32 | 6.48 | -0.011 | 0.026 |
| $Fe_8N$ | 0.111 | $I4/mmm$ | 139 | 5.68 | 5.68 | 6.22 | -0.029 | 0.004 |
| | 0.111 | $P\bar{1}$ | 2 | 4.21 | 4.72 | 5.06 | -0.003 | 0.030 |
| | 0.111 | $C2/m$ | 12 | 5.08 | 8.01 | 5.06 | -0.022 | 0.011 |
| | 0.111 | $P1$ | 1 | 4.73 | 5.06 | 8.76 | -0.017 | 0.016 |
| $Fe_9N$ | 0.1 | $C2/m$ | 12 | 8.54 | 3.98 | 7.56 | -0.016 | 0.014 |
| | 0.1 | $P\bar{1}$ | 2 | 4.20 | 4.71 | 6.17 | -0.008 | 0.021 |
| $Fe_{11}N$ | 0.083 | $P\bar{1}$ | 2 | 4.73 | 4.73 | 6.39 | -0.015 | 0.010 |
| | 0.083 | $C2/m$ | 12 | 12.17 | 3.97 | 6.18 | -0.011 | 0.014 |
| $Fe_{12}N$ | 0.077 | $C2/m$ | 12 | 8.00 | 6.11 | 6.31 | -0.012 | 0.011 |
| | 0.077 | $P\bar{1}$ | 2 | 5.02 | 5.02 | 6.38 | -0.019 | 0.004 |
| $Fe_{13}N$ | 0.071 | $P\bar{1}$ | 2 | 4.72 | 4.96 | 7.00 | -0.004 | 0.017 |
| $Fe_{14}N$ | 0.067 | $P\bar{1}$ | 2 | 5.01 | 6.34 | 6.35 | -0.013 | 0.007 |
| $Fe_{15}N$ | 0.062 | $C2/m$ | 12 | 12.01 | 3.99 | 7.78 | -0.007 | 0.012 |
| | 0.062 | $P\bar{1}$ | 2 | 4.72 | 6.34 | 6.42 | -0.010 | 0.009 |
| $Fe_{16}N$ | 0.059 | $Fmmm$ | 69 | 5.66 | 11.32 | 12.00 | -0.007 | 0.011 |
| $Fe_{19}N$ | 0.05 | $P\bar{1}$ | 2 | 6.19 | 6.32 | 6.99 | -0.007 | 0.008 |
| $Fe_{20}N$ | 0.048 | $P\bar{1}$ | 2 | 6.31 | 6.33 | 6.57 | -0.004 | 0.010 |
| Fe | 0 | $Im\bar{3}m$ | 229 | 2.83 | 2.83 | 2.83 | 0.000 | 0.000 |

**Table A2:** A summary of the phases, compositions, space groups, bulk moduli ($B_H$), shear moduli ($G_H$), Young's moduli ($E$), Poisson's ratios ($v$), Pugh's ratios ($k$), Cauchy pressures ($P_C$), Kleinman's parameters ($\zeta$), universal elastic anisotropies ($A^U$), and Debye temperatures ($\Theta_D$) of stable and metastable Fe-N compounds predicted by USPEX.

| Phase | Composition N/(Fe+N) | Space group | $B_H$ (GPa) | $G_H$ (GPa) | $E$ (GPa) | $v$ | $k$ | $P_C$ (GPa) | $\zeta$ | $A^U$ | $\Theta_D$ (K) |
|---|---|---|---|---|---|---|---|---|---|---|---|
| FeN | 0.5 | $F\bar{4}3m$ | 310.33 | 95.93 | 260.90 | 0.36 | 3.24 | 162.7 | 0.66 | 0.58 | 625.7 |
| $Fe_2N$ | 0.333 | *Pbcn* | 268.97 | 103.27 | 274.65 | 0.33 | 2.61 | 122.7 | 0.70 | 0.17 | 589.6 |
| | 0.333 | *Pnnm* | 265.90 | 100.54 | 267.85 | 0.33 | 2.65 | 146.6 | 0.68 | 0.33 | 581.6 |
| $Fe_7N_3$ | 0.3 | $P6_3$ | 251.33 | 101.65 | 268.72 | 0.32 | 2.47 | 56.9 | 0.59 | 0.12 | 578.1 |
| $Fe_8N_3$ | 0.273 | *P*1 | 227.75 | 107.97 | 279.72 | 0.30 | 2.11 | 81.1 | 0.63 | 0.29 | 586.5 |
| | 0.273 | *C*2/*m* | 202.73 | 91.09 | 237.66 | 0.31 | 2.23 | 36.4 | 0.54 | 0.09 | 541.1 |
| $Fe_3N$ | 0.25 | $P6_322$ | 203.43 | 93.88 | 244.09 | 0.30 | 2.17 | 40.1 | 0.76 | 0.04 | 546.2 |
| | 0.25 | *Pm* | 209.90 | 80.12 | 213.23 | 0.33 | 2.62 | 51.6 | 0.65 | 0.68 | 505.3 |
| | 0.25 | *C*2/*m* | 199.74 | 84.60 | 222.41 | 0.31 | 2.36 | 44.2 | 0.56 | 0.15 | 519.4 |
| | 0.25 | *Immm* | 214.37 | 76.68 | 205.54 | 0.34 | 2.80 | 34.6 | 0.51 | 1.67 | 494.8 |
| $Fe_{10}N_3$ | 0.231 | *P*1 | 202.22 | 72.91 | 195.25 | 0.34 | 2.77 | 63.1 | 0.60 | 0.75 | 480.4 |
| $Fe_{11}N_3$ | 0.214 | $P\bar{1}$ | 181.20 | 86.96 | 224.91 | 0.29 | 2.08 | 38.2 | 0.64 | 0.57 | 520.5 |
| $Fe_4N$ | 0.2 | $Pm\bar{3}m$ | 190.72 | 48.28 | 133.57 | 0.38 | 3.95 | 102.1 | 0.58 | 1.13 | 391.0 |
| | 0.2 | *I*4/*mmm* | 158.29 | 62.20 | 164.98 | 0.33 | 2.55 | 59.8 | 0.55 | 0.77 | 438.1 |
| | 0.2 | *Pmna* | 190.14 | 62.15 | 168.13 | 0.35 | 3.06 | 85.6 | 0.72 | 2.60 | 438.4 |
| | 0.2 | $P\bar{1}$ | 182.90 | 78.63 | 206.33 | 0.31 | 2.33 | 55.3 | 0.55 | 0.47 | 491.6 |
| | 0.2 | *P*1 | 171.84 | 63.23 | 168.97 | 0.34 | 2.72 | 43.5 | 0.61 | 2.65 | 441.8 |
| | 0.2 | *Pbcn* | 142.92 | 44.23 | 120.27 | 0.36 | 3.23 | 109.0 | 0.53 | 10.11 | 370.7 |
| | 0.2 | *Fmmm* | 181.37 | 56.10 | 152.58 | 0.36 | 3.23 | 57.4 | 0.65 | 3.07 | 418.0 |
| | 0.2 | *C*2 | 201.07 | 79.53 | 210.80 | 0.33 | 2.53 | 50.7 | 0.51 | 0.32 | 493.9 |
| | 0.2 | *Cmcm* | 173.49 | 94.69 | 240.34 | 0.27 | 1.83 | -10.6 | 0.44 | 0.23 | 536.2 |
| | 0.2 | *Fdd*2 | 146.09 | 67.84 | 176.23 | 0.30 | 2.15 | 67.6 | 0.60 | 3.01 | 455.4 |
| $Fe_9N_2$ | 0.182 | *C*2 | 192.27 | 83.82 | 219.56 | 0.31 | 2.29 | 61.5 | 0.64 | 0.30 | 504.8 |
| | 0.182 | *C*2/*m* | 182.09 | 54.18 | 147.88 | 0.37 | 3.36 | 38.4 | 0.54 | 1.22 | 408.4 |
| $Fe_5N$ | 0.167 | *C*2/*m* | 191.91 | 78.29 | 206.75 | 0.32 | 2.45 | 58.3 | 0.65 | 0.54 | 486.2 |
| | 0.167 | *C*2 | 191.65 | 73.01 | 194.35 | 0.33 | 2.63 | 92.3 | 0.62 | 0.91 | 469.6 |
| $Fe_{16}N_3$ | 0.158 | $P\bar{1}$ | 191.08 | 86.86 | 226.29 | 0.30 | 2.20 | 47.7 | 0.55 | 0.48 | 508.6 |
| $Fe_6N$ | 0.143 | *P*1 | 198.72 | 86.66 | 226.99 | 0.31 | 2.29 | 66.9 | 0.59 | 0.37 | 505.7 |
| | 0.143 | *C*2/*m* | 197.94 | 76.99 | 204.47 | 0.33 | 2.57 | 70.0 | 0.68 | 0.66 | 478.5 |
| $Fe_{13}N_2$ | 0.133 | *C*2 | 190.11 | 84.39 | 220.53 | 0.31 | 2.25 | 54.4 | 0.55 | 0.40 | 497.9 |
| $Fe_7N$ | 0.125 | *C*2/*m* | 194.70 | 89.78 | 233.45 | 0.30 | 2.17 | 6.6 | 0.47 | 0.31 | 511.8 |
| | 0.125 | *P*1 | 186.15 | 87.42 | 226.75 | 0.30 | 2.13 | 34.4 | 0.49 | 0.39 | 505.1 |
| | 0.125 | $P\bar{1}$ | 188.08 | 89.09 | 230.83 | 0.30 | 2.11 | 56.1 | 0.56 | 0.39 | 509.6 |
| $Fe_8N$ | 0.111 | *I*4/*mmm* | 197.87 | 94.08 | 243.62 | 0.30 | 2.10 | 36.0 | 0.60 | 0.16 | 521.3 |
| | 0.111 | $P\bar{1}$ | 187.80 | 88.67 | 229.83 | 0.30 | 2.12 | 60.2 | 0.54 | 0.40 | 506.0 |
| | 0.111 | *C*2/*m* | 190.14 | 93.94 | 241.97 | 0.29 | 2.02 | 32.4 | 0.52 | 0.21 | 520.7 |
| | 0.111 | *P*1 | 189.82 | 92.03 | 237.68 | 0.29 | 2.06 | 38.3 | 0.54 | 0.34 | 515.1 |
| $Fe_9N$ | 0.1 | *C*2/*m* | 190.33 | 92.22 | 238.18 | 0.29 | 2.06 | -4.2 | 0.43 | 0.32 | 513.8 |
| | 0.1 | $P\bar{1}$ | 187.36 | 92.01 | 237.20 | 0.29 | 2.04 | 23.4 | 0.50 | 0.31 | 513.2 |
| $Fe_{11}N$ | 0.083 | $P\bar{1}$ | 190.26 | 94.83 | 243.97 | 0.29 | 2.01 | 46.5 | 0.54 | 0.32 | 518.1 |
| | 0.083 | *C*2/*m* | 185.20 | 91.78 | 236.31 | 0.29 | 2.02 | 23.6 | 0.50 | 0.35 | 509.4 |
| $Fe_{12}N$ | 0.077 | *C*2/*m* | 196.51 | 91.95 | 238.63 | 0.30 | 2.14 | 29.9 | 0.52 | 0.25 | 510.3 |
| | 0.077 | $P\bar{1}$ | 193.82 | 93.56 | 241.77 | 0.29 | 2.07 | 13.4 | 0.47 | 0.33 | 514.0 |
| $Fe_{13}N$ | 0.071 | $P\bar{1}$ | 183.50 | 91.13 | 234.56 | 0.29 | 2.01 | 20.9 | 0.50 | 0.36 | 505.9 |
| $Fe_{14}N$ | 0.067 | $P\bar{1}$ | 192.90 | 94.46 | 243.62 | 0.29 | 2.04 | 58.2 | 0.58 | 0.32 | 514.2 |
| $Fe_{15}N$ | 0.062 | *C*2/*m* | 178.93 | 90.76 | 232.91 | 0.28 | 1.97 | -21.9 | 0.42 | 0.26 | 502.9 |
| | 0.062 | $P\bar{1}$ | 191.02 | 93.99 | 242.25 | 0.29 | 2.03 | 25.8 | 0.54 | 0.34 | 512.5 |
| $Fe_{16}N$ | 0.059 | *Fmmm* | 192.71 | 88.69 | 230.68 | 0.30 | 2.17 | 51.2 | 0.62 | 0.37 | 498.1 |
| $Fe_{19}N$ | 0.05 | $P\bar{1}$ | 180.71 | 88.17 | 227.50 | 0.29 | 2.05 | 38.3 | 0.51 | 0.33 | 494.3 |
| $Fe_{20}N$ | 0.048 | $P\bar{1}$ | 183.81 | 87.49 | 226.53 | 0.30 | 2.10 | 0.6 | 0.47 | 0.30 | 492.4 |
| Fe | 0 | $Im\bar{3}m$ | 175.18 | 77.54 | 202.71 | 0.31 | 2.26 | 45.8 | 0.77 | 0.01 | 457.0 |

**Table A3:** A summary of the phases, compositions, space groups, Fe-N bond lengths (*d*), overlap populations (*p*), total DOS values at the Fermi level ($D_F$), the number of the valence electrons per cell ($N_e$), volumes of primitive cells (*V*), and Vickers hardness ($V_H$) of stable and metastable Fe-N compounds predicted by USPEX.

| Phase | Composition N/(Fe+N) | Space group | *d* (Å) | *p* | $D_F$ | $N_e$ | *V* (Å$^3$) | $V_H$ (GPa) |
|---|---|---|---|---|---|---|---|---|
| FeN | 0.5 | $F\bar{4}3m$ | 1.82 | 1.47 | 1.26 | 13 | 18.71 | 10.17 |
| $Fe_2N$ | 0.333 | *Pbcn* | 1.91 | 0.25 | 12.65 | 84 | 111.38 | 4.28 |
| | 0.333 | *Pnnm* | 1.91 | 0.34 | 5.85 | 42 | 56.09 | 5.32 |
| $Fe_7N_3$ | 0.3 | $P6_3$ | 1.92 | 0.27 | 26.34 | 142 | 191.99 | 3.59 |
| $Fe_8N_3$ | 0.273 | *P*1 | 1.94 | 0.27 | 11.74 | 79 | 106.01 | 4.23 |
| | 0.273 | *C*2/*m* | 1.93 | 0.27 | 11.36 | 79 | 108.10 | 4.35 |
| $Fe_3N$ | 0.25 | $P6_322$ | 1.90 | 0.22 | 5.45 | 58 | 80.94 | 5.68 |
| | 0.25 | *Pm* | 1.91 | 0.37 | 6.33 | 58 | 80.00 | 6.61 |
| | 0.25 | *C*2/*m* | 1.91 | 0.31 | 8.57 | 58 | 81.18 | 4.62 |
| | 0.25 | *Immm* | 1.91 | 0.44 | 3.26 | 29 | 39.94 | 7.44 |
| $Fe_{10}N_3$ | 0.231 | *P*1 | 1.91 | 0.27 | 12.59 | 95 | 133.43 | 4.70 |
| $Fe_{11}N_3$ | 0.214 | $P\bar{1}$ | 1.89 | 0.31 | 12.84 | 103 | 148.44 | 5.36 |
| $Fe_4N$ | 0.2 | $Pm\bar{3}m$ | 1.89 | 0.56 | 3.84 | 37 | 53.85 | 9.94 |
| | 0.2 | *I*4/*mmm* | 1.90 | 0.30 | 5.44 | 37 | 52.15 | 4.64 |
| | 0.2 | *Pmna* | 1.89 | 0.24 | 12.81 | 74 | 102.85 | 3.71 |
| | 0.2 | $P\bar{1}$ | 1.91 | 0.27 | 15.47 | 111 | 156.39 | 4.53 |
| | 0.2 | *P*1 | 1.90 | 0.24 | 9.87 | 74 | 103.64 | 4.58 |
| | 0.2 | *Pbcn* | 1.89 | 0.27 | 22.13 | 148 | 207.32 | 4.38 |
| | 0.2 | *Fmmm* | 1.93 | 0.25 | 9.42 | 74 | 104.34 | 4.60 |
| | 0.2 | *C*2 | 1.89 | 0.24 | 10.43 | 74 | 102.48 | 4.44 |
| | 0.2 | *Cmcm* | 1.93 | 0.36 | 10.41 | 74 | 103.79 | 5.12 |
| | 0.2 | *Fdd*2 | 1.89 | 0.26 | 7.42 | 74 | 103.60 | 5.84 |
| $Fe_9N_2$ | 0.182 | *C*2 | 1.91 | 0.25 | 8.64 | 82 | 116.53 | 5.34 |
| | 0.182 | *C*2/*m* | 1.90 | 0.31 | 14.22 | 82 | 115.84 | 4.08 |
| $Fe_5N$ | 0.167 | *C*2/*m* | 1.91 | 0.25 | 4.55 | 45 | 64.27 | 5.51 |
| | 0.167 | *C*2 | 1.91 | 0.26 | 9.92 | 90 | 127.60 | 5.32 |
| $Fe_{16}N_3$ | 0.158 | $P\bar{1}$ | 1.92 | 0.30 | 15.15 | 143 | 202.30 | 5.77 |
| $Fe_6N$ | 0.143 | *P*1 | 1.90 | 0.28 | 12.53 | 106 | 150.01 | 5.30 |
| | 0.143 | *C*2/*m* | 1.91 | 0.35 | 4.92 | 53 | 75.68 | 6.94 |
| $Fe_{13}N_2$ | 0.133 | *C*2 | 1.91 | 0.27 | 13.06 | 114 | 162.78 | 5.23 |
| $Fe_7N$ | 0.125 | *C*2/*m* | 1.90 | 0.31 | 5.54 | 61 | 87.26 | 6.61 |
| | 0.125 | *P*1 | 1.91 | 0.28 | 13.28 | 122 | 174.97 | 5.48 |
| | 0.125 | $P\bar{1}$ | 1.92 | 0.29 | 12.10 | 122 | 174.56 | 5.83 |
| $Fe_8N$ | 0.111 | *I*4/*mmm* | 1.90 | 0.27 | 7.31 | 69 | 98.99 | 5.53 |
| | 0.111 | $P\bar{1}$ | 1.91 | 0.25 | 6.02 | 69 | 98.78 | 6.04 |
| | 0.111 | *C*2/*m* | 1.91 | 0.24 | 5.93 | 69 | 99.23 | 6.00 |
| | 0.111 | *P*1 | 1.91 | 0.27 | 13.78 | 138 | 197.22 | 5.76 |
| $Fe_9N$ | 0.1 | *C*2/*m* | 1.90 | 0.33 | 6.99 | 77 | 110.21 | 6.86 |
| | 0.1 | $P\bar{1}$ | 1.90 | 0.25 | 6.88 | 77 | 110.41 | 6.01 |
| $Fe_{11}N$ | 0.083 | $P\bar{1}$ | 1.91 | 0.27 | 7.81 | 93 | 133.59 | 6.37 |
| | 0.083 | *C*2/*m* | 1.90 | 0.37 | 10.33 | 93 | 133.12 | 6.44 |
| $Fe_{12}N$ | 0.077 | *C*2/*m* | 1.90 | 0.27 | 7.52 | 101 | 146.04 | 6.89 |
| | 0.077 | $P\bar{1}$ | 1.90 | 0.28 | 8.11 | 101 | 145.38 | 6.69 |
| $Fe_{13}N$ | 0.071 | $P\bar{1}$ | 1.91 | 0.27 | 9.60 | 109 | 156.66 | 6.22 |
| $Fe_{14}N$ | 0.067 | $P\bar{1}$ | 1.91 | 0.29 | 8.64 | 117 | 167.69 | 7.15 |
| $Fe_{15}N$ | 0.062 | *C*2/*m* | 1.92 | 0.37 | 12.43 | 125 | 179.11 | 6.80 |
| | 0.062 | $P\bar{1}$ | 1.91 | 0.28 | 8.40 | 125 | 179.76 | 7.40 |
| $Fe_{16}N$ | 0.059 | *Fmmm* | 1.91 | 0.31 | 8.63 | 133 | 191.70 | 7.89 |
| $Fe_{19}N$ | 0.05 | $P\bar{1}$ | 1.91 | 0.30 | 14.28 | 157 | 225.36 | 6.40 |
| $Fe_{20}N$ | 0.048 | $P\bar{1}$ | 1.91 | 0.33 | 15.29 | 165 | 237.18 | 6.68 |